\documentclass[fleqn,twoside]{article}
\usepackage{espcrc2}
\usepackage{epsfig}
\usepackage{graphicx}
\def\lsim{\buildrel < \over {_{\sim}}}

\newcommand{\magq}{|{\bf q}|}

\newcommand{\magp}{|{\bf p}|}

\newcommand{\vecq}{{\bf q}}
\newcommand{\vecp}{{\bf p}}
\newcommand{\beq}{\begin{equation}}
\newcommand{\eeq}{\end{equation}}
\newcommand{\be}{\begin{eqnarray}}
\newcommand{\ee}{\end{eqnarray}}
\title{Nuclear response beyond the Fermi gas model}

\author{Omar Benhar \\ ~~~ \\
INFN, Sezione di Roma, and Department of Physics, 
Universit\`a ``La Sapienza'' \\
        Piazzale Aldo Moro, 2. I-00185 Roma, Italy}
       
\begin{document}

\begin{abstract}
 The Fermi gas model, while providing a reasonable qualitative 
description of the continuum nuclear response, does not include
the effects of dynamical nucleon-nucleon correlations in 
the initial and final states, that have long been recognized
to play a critical role in specific kinematical regions. We 
review a many-body approach in which these effects are consistently 
taken into account and discuss the 
results of a calculation of the quasielastic neutrino-oxygen cross 
section as an illustrative example.
\vspace{1pc}
\end{abstract}

\maketitle

\section{Introduction}

Within the Fermi gas (FG) model the nucleus is seen as a degenerate
gas of protons and neutrons at density $\rho$, binding effects being 
taken into account
through the inclusion of 
an average interaction energy ${\overline \epsilon}$ \cite{Moniz1}. 

In spite of its simplicity, the FG model provides a remarkably good 
description of 
the inclusive electron-nucleus cross section in the region of the 
quasi free peak, corresponding to $x_B = Q^2/2m\nu \sim 1$, where 
$x_B$ is the Bjorken scaling variable, $Q^2 = {\bf q}^2 - \nu^2$, 
${\bf q}$ and $\nu$ being the momentum and energy transfer, respectively, 
and $m$ denotes the nucleon mass. The successful application 
to the analysis of electron scattering data at moderate momentum transfer
(${\bf q} \sim 150 - 200$ MeV) \cite{Moniz2} also prompted 
the extension of the FG approach to the case of neutrino-nucleus scattering 
\cite{Moniz3}.

\begin{figure}[hbt]
\centerline{\psfig{figure=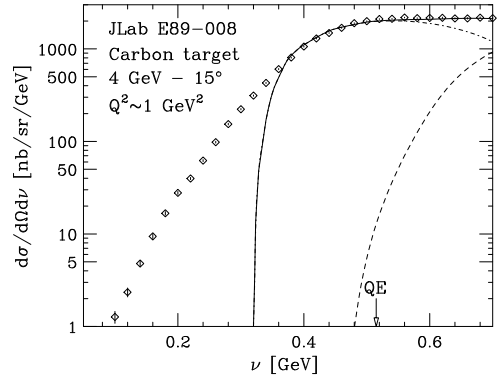,angle=00,height=5.0cm}}
\vspace*{-.2in}
\caption{\small Cross section versus energy loss for scattering of 4 GeV 
electrons at 15 $^\circ$ from Carbon. The dot-dash, dashed and full lines 
show the quasielastic, inelastic and total cross sections predicted by  
the FG model with $k_F = 221$ MeV and ${\overline \epsilon} = 35$ MeV. 
The arrow points to the value of $\nu$ corresponding to quasi-free kinematics.
The experimental data are taken from ref. \cite{Arrington}.}
\label{fig:1}
\end{figure}

Over the past two decades the limitations of the FG description have been 
exposed by the growing availability of new electron 
scattering data, extending into the region of large ${\bf q}$ and low 
$\nu$, corresponding to $x_B \gg 1$. The failure of the FG model 
to explain the measured cross sections 
in this kinematical regime, clearly
illustrated in Fig. \ref{fig:1}, has been ascribed to the dominance
of scattering off nucleons belonging to {\it strongly correlated} pairs, 
that cannot be accounted for by any mean field approach.

In addition, it has to be pointed out that, as the FG model does not 
include surface and shell effects, it cannot be employed to study 
exclusive channels, which are known to be sensitive to the details of 
the nuclear wave function.
%%%%%%%%%%%%%%%%%%%%%%%%%%%%%%%%%%%%%%%%%%%%%%%%%%%%%%%%%%%%%%%%%%%%%%%%%%%%
%%%%%%%%%%%%%%%%%%%%%%%%%%%%%%%%%%%%%%%%%%%%%%%%%%%%%%%%%%%%%%%%%%%%%%%%%%%%%%

In this paper we review a theoretical approach, based on 
nonrelativistic nuclear many-body theory, that incorporates in a
consistent fashion dynamical nucleon-nucleon (NN) correlations in both the 
initial and final state.
The impulse approximation (IA) scheme, which allows one to relate
the cross section to the nuclear spectral function $P({\bf p},E)$,
describing the energy-momentum distributions of nucleons in the target nucleus,
is outlined in Section 2. The main features of $P({\bf p},E)$, extracted
from both experiments and theoretical calculations, are described in
Section 3, while Section 4 is devoted to the discussion of final state
interactions (FSI). Finally, in Section 5 we summarize the 
 results and state the conclusions.

\section{The impulse approximation}

The main assumption underlying IA is that, as
the space resolution of a probe delivering momentum $\vecq$ to a nucleus
is $\sim 1/\magq$, at large enough $\magq$
the target is seen by the probe as a collection of individual nucleons.
Within this picture, the nuclear response measures the probability that, 
after giving one nucleon a momentum $\vecq$ at time $t=0$, the 
system be reverted to the ground state giving the {\it same} 
particle a momentum $-\vecq$ after time $t$.

The second assumption involved in IA is that (FSI) taking place 
at $0<t^\prime<t$
between the hit nucleon and the (A-1)-nucleon spectator system be
negligible.

\begin{figure}[hbt]
\centerline{\psfig{figure=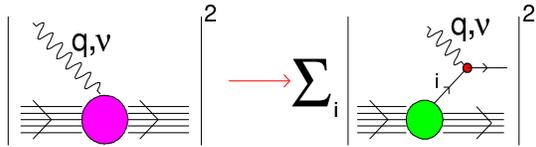
,angle=00,width=7.0cm,height=2.00cm}}
\vspace*{-.2in}
\caption{\small Pictorial representation of the IA scheme, in which
the nuclear cross section is replaced by the incoherent sum of
cross sections describing scattering off individual nucleons, the
recoiling (A-1)-particle system acting as a spectator.}
\label{fig:2}
\end{figure}

In the IA regime the scattering process off a nuclear target reduces to the
incoherent sum of elementary processes involving only one nucleon, as 
schematically illustrated in Fig. \ref{fig:2}.  
%%%%%%%%%%%%%%%%%%%%%%%%%%%%%%%%%%%%%%%%%%%%%%%%%%%%%%%%%%%%%%%%%%%%%%%%%%%%
%%%%%%%%%%%%%%%%%%%%%%%%%%%%%%%%%%%%%%%%%%%%%%%%%%%%%%%%%%%%%%%%%%%%%%%%%%%%%%

Let us consider the weak charged 
current reaction
\beq
\nu_\mu + A \rightarrow \mu^- + p + (A-1)\ .
\eeq
According to IA, the corresponding differential cross section can be written 
\be
\nonumber
\frac{d\sigma_{IA}}{d\Omega_\mu dE_\mu} & = &
\int d^3p\ dE\ P({\bf p},E) \frac{d\sigma_N}{d\Omega_\mu dE_\mu} \\
& & \ \ \ \ \times \delta(E_\nu-E_\mu-E-E_{{\bf p}^\prime} )\ ,
\label{sigma:IA}
\ee
where $d\sigma_N/d\Omega_\mu dE_\mu$ is the cross section describing the 
elementary scattering
process
\beq
\nu_\mu(k) + n(p) \rightarrow \mu^-(k^\prime) + p(p^\prime)\ ,
\eeq
involving a {\it bound} nucleon {\it carrying momentum} $\vecp$. 
The spectral function $P({\bf p},E)$, yielding the probability
of removing a nucleon with momentum $\vecp$ from the nuclear ground state
leaving the residual system with excitation energy $E$, will be 
discussed in Section 3.

Up to a kinematical factor, the elementary cross section reads
\beq
\frac{d\sigma_N}{d\Omega_\mu dE_\mu} \propto \frac{G^2}{2} 
L_{\mu \nu}(k,k^\prime) W^{\mu \nu}(p,{\widetilde q})\ ,
\eeq
where $G=G_F\cos\theta_c$, $G_F$ and $\theta_c$ being Fermi's coupling 
constant and 
Cabibbo's angle, respectively.
The lepton tensor $L_{\mu \nu}$ is totally specified by 
kinematical variables, whereas the definition of 
$W^{\mu \nu}(p,{\widetilde q})$, describing
the weak interactions of a {\it free} nucleon, involves the
Dirac, Pauli and axial form factors $F_1$, $F_2$ and $F_A$. 
%%%%%%%%%%%%%%%%%%%%%%%%%%%%%%%%%%%%%%%%%%%%%%%%%%%%%%%%%%%%%%%%%%%%%%%%%%%%
\begin{figure}[hbt]
\centerline{\psfig{figure=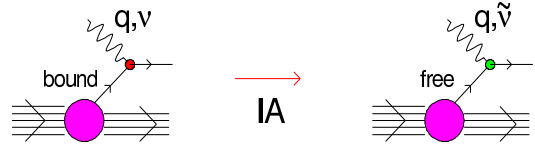
,angle=00,width=7.0cm,height=2.00cm}}
\vspace*{-.2in} 
\caption{\small Schematic representation of the IA treatment of 
the lepton nucleon vertex. The elementary scattering cross section is 
assumed to be
the same as in free space, binding effects being taken into account 
replacing the energy transfer $\nu$ with ${\widetilde \nu}$ (see text).} 
\label{fig:3}
\end{figure}
%%%%%%%%%%%%%%%%%%%%%%%%%%%%%%%%%%%%%%%%%%%%%%%%%%%%%%%%%%%%%%%%%%%%%%%%%%%%%%

Within the
IA scheme binding is taken into account replacing the physical four-momentum
transfer $q=k-k^\prime \equiv(\nu,{\bf q})$ with 
${\widetilde q}\equiv({\widetilde \nu},{\bf q})$, where
\beq
{\widetilde \nu} = \nu + M_A - \sqrt{\magp^2 + M_{A-1}^2} - 
\sqrt{\magp^2 + m^2}\ , 
\eeq 
$M_A$ and $M_{A-1} = M_A -m + E$ being the target mass and the mass of the 
recoiling
nucleus, respectively \cite{defo}. This essentially amounts to assuming 
that: i) a fraction 
$(\nu - {\widetilde \nu})/\nu$ of the lepton energy loss is spent to put the 
struck nucleon on the mass shell and ii) the elementary
scattering process can be described as if it took place in free space 
with energy transfer ${\widetilde \nu}<\nu$. 

\section{The nuclear spectral function}

Let us consider a system of A nucleons whose dynamics is described by
the nonrelativistic hamiltonian
\beq
H_A = \sum_{i=1}^{A} \frac{{\bf p}_i^2}{2m} + \sum_{j>i=1}^{A} v_{ij}
 + \ldots \ ,
\label{H:A}
\eeq
where ${\bf p}_i$ is the momentum of the $i$-th nucleon, $v_{ij}$ is the 
 NN potential and the ellipsis 
indicates the presence of three-nucleon interactions 
\footnote{The inclusion of a three-nucleon potential in nuclear 
many-body theory is needed to reproduce the binding energy of the 
three-nucleon
system and the empirical equilibrium density of nuclear matter.}.

The spectral function is defined as (see, e.g., ref. \cite{bff})
\be
\nonumber
P({\bf p},E)&=&\sum_{n} \left|
\langle \Psi_n^{(A-1)} | a_{{\bf p}} | \Psi_0^{A} \rangle \right|^2 \\
 & & \ \ \ \ \ \ \ \ \ \ \ \ \ \times \delta( E + E_0 - E_n )\ , 
\label{def:pke}
\ee
where $| \Psi_0^{A}\rangle$ is the nuclear ground state, satisfying the 
Scr\"odinger equation $H_A | \Psi_0^{A}\rangle = E_0| \Psi_0^{A}\rangle$, 
while $|\Psi_n^{(A-1)}\rangle$ and $E_n$ denote the $n$-th eigenstate of the 
(A-1)-nucleon system and the associated energy eigenvalue, respectively.

Due to the complex spectrum of (A-1)-nucleon states, in general the 
calculation of $P({\bf p},E)$ within nuclear many-body theory involves 
prohibitive difficulties. It has been carried out only for few-nucleon 
systems,
having A$=3$ \cite{cps,sauer} and 4 \cite{bp}, and uniform symmetric 
nuclear matter, i.e. in the limit A $\rightarrow \infty$ with Z=A/2
\cite{bff,pkebbg}. 

Spectral functions of nuclei with A $>$ 4 have been constructed using the
Local Density Approximation (LDA) \cite{bffs}.
%, in which the experimental
%information obtained from nucleon knock-out measurements is combined 
%with the results of theoretical calculations of the nuclear matter 
%$P({\bf p},E)$ at different densities \cite{bffs}.
Within this approach $P({\bf p},E)$ is split into two parts, 
corresponding to low momentum nucleons, sitting in single particle 
shell model states, and high-momentum nucleons, belonging to strongly 
correlated pairs, respectively. 

The spectral function of the shell model states has been extensively 
measured by single nucleon knock-out experiments (see, e.g., ref. \cite{book}), 
whereas the correlation
part is obtained from the results of theoretical calculations 
through \cite{bffs}
\beq
P_{corr}({\bf p},E) = \int d^3r\ \rho({\bf r}) 
P^{NM}_{corr}({\bf p},E;\rho({\bf r}))
\eeq
where $\rho({\bf r})$ is the nuclear density distribution and 
$P^{NM}_{corr}({\bf p},E;\rho)$ is the correlation part of
the spectral function of infinite nuclear matter at uniform density $\rho$.
%%%%%%%%%%%%%%%%%%%%%%%%%%%%%%%%%%%%%%%%%%%%%%%%%%%%%%%%%%%%%%%%%%%%%%%%%%%%
\begin{figure}[hbt]
\centerline{\psfig{figure=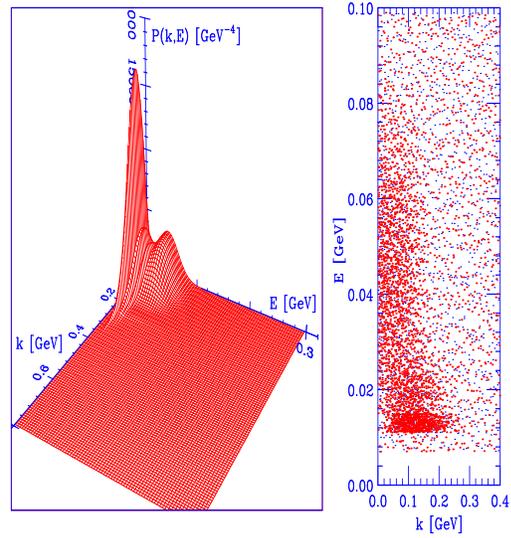
,angle=00,width=6.7cm,height=7.20cm}}
\vspace*{-.2in}
\caption{\small Three-dimensional plot (left panel) and scatter plot 
(right panel) of the oxygen spectral function obtained using the LDA
approximation described in the text.}
\label{pke:O}
\end{figure}
%|%%%%%%%%%%%%%%%%%%%%%%%%%%%%%%%%%%%%%%%%%%%%%%%%%%%%%%%%%%%%%%%%%%%%%%%%%%%%%

The LDA spectral function of $^{16}O$ obtained combining the nuclear matter 
results of ref. \cite{bff} and the Saclay $(e,e^\prime p)$ data \cite{eep16O}
is shown in Fig. \ref{pke:O}. The shell model states account for $\sim$ 80 \%
of the strength, whereas the remaining $\sim$ 20 \% is located at high momentum
($|{\bf p}| \gg k_F$) {\it and} large removal energy 
($E \gg E_F$, where $E_F$ denotes the Fermi energy). It has to be emphasized 
that large $E$ and large ${\bf p}$ are strongly correlated. For example, 
$\sim$ 50 \% of the strength at $|{\bf p}|$ = 320 MeV is located at 
$E >$ 80 MeV.

\begin{figure}[hbt]
\centerline{\psfig{figure=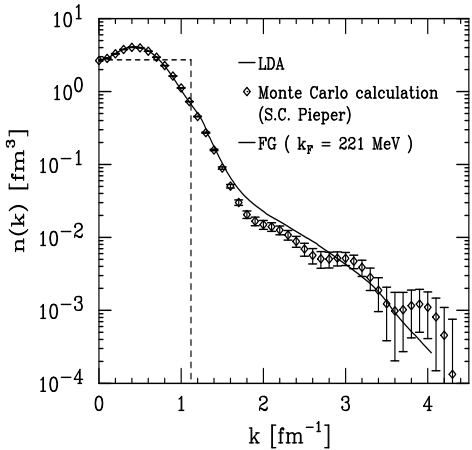
,angle=00,width=6.25cm,height=6.0cm}}
\vspace*{-.2in}
\caption{\small Momentum distribution of nucleons in the oxygen ground state.
Solid line: LDA approximation. Dashed line: Fermi gas model with 
$k_F = 220$ MeV. 
Diamonds: Monte Carlo calculation carried out by S.C. Pieper using the 
wave function of ref. \protect\cite{16Owf}.}
\label{nk:O}
\end{figure}

Fig, \ref{nk:O} shows the momentum distribution of nucleons in the 
ground state of $^{16}$O, obtained from the LDA spectral function
through
\beq
n({\bf p}) = \int dE\ P({\bf p},E)\ .
\label{def:nk}
\eeq
Comparison with the results of a Monte Carlo calculation carried
out using a highly realistic many-body wave 
function \cite{16Owf} suggests that LDA results are quite reasonable. 
For reference, the Fermi gas momemtum distribution corresponding to 
$k_F$ = 220 MeV is also shown by the dashed line.
%%%%%%%%%%%%%%%%%%%%%%%%%%%%%%%%%%%%%%%%%%%%%%%%%%%%%%%%%%%%%%%%%%%%%%%%%%%%
%%%%%%%%%%%%%%%%%%%%%%%%%%%%%%%%%%%%%%%%%%%%%%%%%%%%%%%%%%%%%%%%%%%%%%%%%%%%%%

\begin{figure}[hbt]
\centerline{\psfig{figure=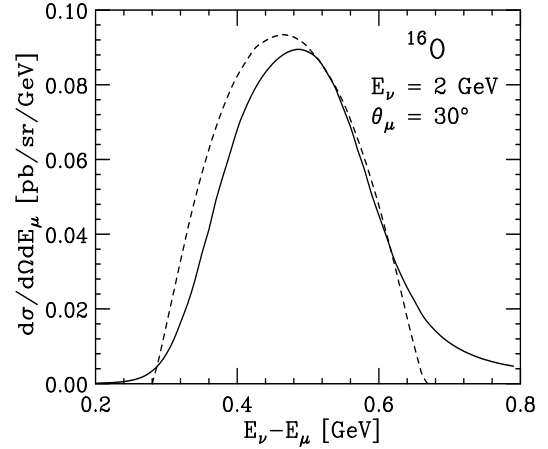
,angle=00,width=7.0cm,height=6.0cm}}
\vspace*{-.2in}
\caption{\small Cross section of the process $\nu_\mu + ^{16}O \rightarrow 
\mu^- + p + X$. Solid line: IA calculation carried out using the LDA spectral
function discussed in the text. Dashed line: Fermi gas model with $k_F =$ 220 
MeV and $\overline \epsilon$ = 25 MeV.}
\label{xsec:1}
\end{figure}

The lepton energy loss dependence of the calculated cross section of 
the process
$\nu_\mu + ^{16}O \rightarrow \mu^- + p + X$ is displayed in Fig. \ref{xsec:1} 
for the case of neutrino energy $E_\nu =$ 2 GeV
and muon scattering angle 30$^\circ$. Comparison between the solid line, 
obtained using the LDA spectral function, and the dashed line, corresponding 
to the FG model, shows that the inclusion of dynamical NN correlations produces
a quenching of the peak associated with the appearance of sizeable tails, 
particularly at large lepton energy loss. 
%%%%%%%%%%%%%%%%%%%%%%%%%%%%%%%%%%%%%%%%%%%%%%%%%%%%%%%%%%%%%%%%%%%%%%%%%%%%
%%%%%%%%%%%%%%%%%%%%%%%%%%%%%%%%%%%%%%%%%%%%%%%%%%%%%%%%%%%%%%%%%%%%%%%%%%%%%%

The presence of high-momentum and high-energy components 
in the nuclear wave function 
is also responsible for the shift in the position of the peaks. 
The average removal energy 
\beq
\langle E \rangle = \int d^3p\ dE\ E P({\bf p},E)\ ,
\eeq
calculated using the LDA spectral function of $^{16}$O, $\langle E \rangle =$
42 MeV, is in fact much larger than the binding energy 
$\overline \epsilon$ = 25 MeV of the FG model.

\section{Final state interactions}

The occurrence of FSI leads to the breakdown of the IA picture of the 
scattering process, thus making the reconstruction of its kinematics much 
more difficult. 

A number of theoretical studies of FSI in $(e,e^\prime p)$ have been carried
out within the Distorded Wave Impulse Approximation \cite{DWIA}, in which the 
spectral function is modified to take into account the distortion of the 
struck nucleon wave function produced by its interactions with the 
spectators, described by a complex optical potential. DWIA has been 
succesfully employed to analize two-body breakup processes, in which the
spectator system is left in a bound state, for proton energies $\lsim$ 100 MeV.

Realistic many-body calculations of nuclear structure provide strong
{\it a priori} indication that the PWIA picture is not likely to be 
applicabile even at much lager energies. 
%%%%%%%%%%%%%%%%%%%%%%%%%%%%%%%%%%%%%%%%%%%%%%%%%%%%%%%%%%%%%%%%%%%%%%%%%%%%
\begin{figure}[hbt]
\centerline{\psfig{figure=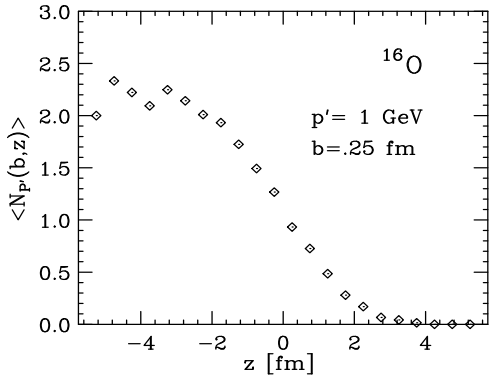
,angle=00,width=6.5cm,height=5.0cm}}
\vspace*{-.2in}
\caption{\small The function $\langle N_{{\bf p}^\prime}({\bf b},z) \rangle$,
defined by Eq.(\protect\ref{rsctn}), plotted at $b =$ 0.25 fm for a 1 GeV
nucleon knocked out from oxygen.}
 \label{nscatt}
\end{figure}
%%%%%%%%%%%%%%%%%%%%%%%%%%%%%%%%%%%%%%%%%%%%%%%%%%%%%%%%%%%%%%%%%%%%%%%%%%%%%%

The probability that a nucleon hit at position ${\bf r}$
in the target center of mass frame and left with momentum ${\bf p}^\prime$
in the direction of the $z$-axis rescatter against the spectator particles
is illustrated in Fig. \ref{nscatt}, showing the $z$-dependence of the
quantity
\[
\langle N_{{\bf p}^\prime}({\bf b},z) \rangle  =   
\frac{1}{\rho({\bf r})} \int dR\ |\Psi_0(R)|^2
\sum_{i=1}^A \delta({\bf r}-{\bf r}_i) 
\]
\beq
\times  \sum_{j \neq i = 1}^A \theta\left( 
\sqrt{\frac{\sigma_{{\rm NN}}({\bf p}^\prime)}{\pi}}
-|{\bf b}- {\bf b}_j|\right) \theta(z_j - z)\ ,
\label{rsctn}
\eeq
evaluated at $|{\bf b}| = \sqrt{ |{\bf r}|^2 - z^2 } = 0.25$ fm for the 
case of a 1 GeV 
nucleon knocked out from oxygen.
In Eq.(\ref{rsctn}) $\rho({\bf r})$ and $\Psi_0(R)$, with 
$R\equiv ({\bf r}_1,\ldots,{\bf r}_A)$, denote the target density and ground 
state wave function, respectively, whereas $\sigma_{NN}({\bf p}^\prime)$ is
the NN scattering cross section at incident momentum 
${\bf p}^\prime$. The 3A-dimensional integral has been carried out using 
Monte Carlo configurations sampled from the probability distribution 
associated with the $^{16}$O wave function of ref. \cite{16Owf}.
%%%%%%%%%%%%%%%%%%%%%%%%%%%%%%%%%%%%%%%%%%%%%%%%%%%%%%%%%%%%%%%%%%%%%%%%%%%%
\begin{figure}[hbt]
\centerline{\psfig{figure=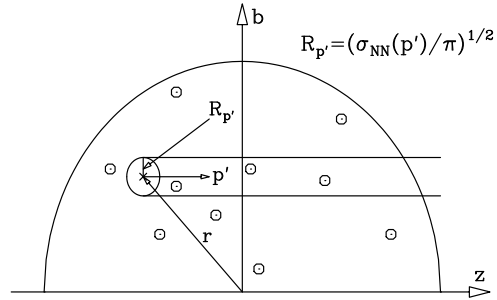
,angle=00,width=6.5cm,height=4.0cm}}
\vspace*{-.2in}
\caption{\small Schematic representation of rescattering in nucleon
knock-out processes.}
\label{tube}
\end{figure}
%%%%%%%%%%%%%%%%%%%%%%%%%%%%%%%%%%%%%%%%%%%%%%%%%%%%%%%%%%%%%%%%%%%%%%%%%%%%%%

The quantity $\langle N_{{\bf p}^\prime}({\bf b},z) \rangle$ is the average
number of spectators found within the cylindrical volume described by a
circle of area $\sigma_{NN}({\bf p}^\prime)$ initially centered at ${\bf r}$
and moving along the direction of the momentum ${\bf p}^\prime$ 
(see Fig. \ref{tube}). Its value 
provides an estimate of the number of rescatterings undergone by the knocked
out particle. The results shown in Fig. \ref{nscatt} suggest that in the 
case of scattering processes involving nucleons in the $z<0$ emisphere, 
the struck particle is likely to interact with at least one of the 
spectators.

A theoretical treatment of the corrections to the IA cross section
due to FSI effects was developed in ref. \cite{gofsix} and extensively applied
to the analysis of inclusive electron-nucleus scattering data 
\cite{bp,bffs,gofsix}. 
The main approximations involved in the approach of ref. \cite{gofsix}, 
expected to be reasonable at large ${\bf p}^\prime$, are that i) the struck 
particle moves 
along a straight trajectory with a constant velocity ${\bf v}$ ({\it eikonal 
approximation}) and ii) the spectator particles are seen by the fast struck 
nucleon as a collection of fixed scattering centers 
({\it frozen approximation}). 

Under the above assumptions the ground state averaged propagator of the
struck nucleon can be written in the factorized form
\beq
U_{{\bf p}^\prime}(t) = U^0_{{\bf p}^\prime}(t) 
U^{FSI}_{{\bf p}^\prime}(t)\ , 
\eeq
where $U^0_{{\bf p}^\prime}$ is the free space propagator, while FSI 
effects are described by the quantity 
\beq
U^{FSI}_{{\bf p}^\prime}(t) = \langle \frac{1}{A} \sum_{i=1}^A
{\rm e}^{i \sum_{j \neq i} \int_0^t dt^\prime 
w_{{\bf p}^\prime}(|{\bf r}_{ij} + {\bf v}t^\prime |) } \rangle\ .
\label{eik:prop}
\eeq
In Eq.(\ref{eik:prop}) ${\bf r}_{ij}={\bf r}_{i}-{\bf r}_{j}$, 
$w_{{\bf p}^\prime}(|{\bf r}|)$ is the coordinate-space NN scattering 
t-matrix, parametrized in terms of total cross section, slope and 
real to imaginary part ratio, and $\langle \ldots \rangle$ denotes
the expectation value in the target ground state. 

Note that 
$U^{FSI}_{{\bf p}^\prime}(t)$ is simply related to the nuclear 
transparency $T_{{\bf p}^\prime}$, 
measured in coincidence $(e,e^\prime p)$ experiments \cite{abbott}, through
\beq
T_{{\bf p}^\prime} = 
\lim_{t \rightarrow \infty} |U^{FSI}_{{\bf p}^\prime}(t)|^2\ .
\eeq

In presence of FSI the inclusive cross section can still be expressed
in terms of the IA result of Eq.(\ref{sigma:IA}) through \cite{gofsix}
\beq
\frac{d\sigma}{d\Omega_\mu dE_\mu} = \int dE_\mu^\prime \ 
\frac{d\sigma_{IA}}{d\Omega_\mu dE_\mu^\prime}\ 
f_{{\bf p}^\prime}(E_\mu - E_\mu^\prime)\ ,
\label{sigma:FSI}
\eeq
the folding function $f_{{\bf p}^\prime}(\omega)$ being defined as
\be
\nonumber
f_{{\bf p}^\prime}(\omega) & = & \delta(\omega) T_{{\bf p}^\prime}^{1/2} \\
& + &  
\int \frac{dt}{2 \pi}\ {\rm e}^{i \omega t} 
\left[ U^{FSI}_{{\bf p}^\prime}(t) - T_{{\bf p}^\prime}^{1/2} \right]\ .
\label{ff}
\ee
The above equations clearly show that the strength of FSI is measured by
both $T_{{\bf p}^\prime}$ and the width of the folding function. In absence 
of FSI
$U^{FSI}_{{\bf p}^\prime}(t) \equiv 1$, implying in turn 
$T_{{\bf p}^\prime}=1$ and $f_{{\bf p}^\prime}(\omega) 
\rightarrow \delta(\omega)$.

\begin{figure}[hbt]
\centerline{\psfig{figure=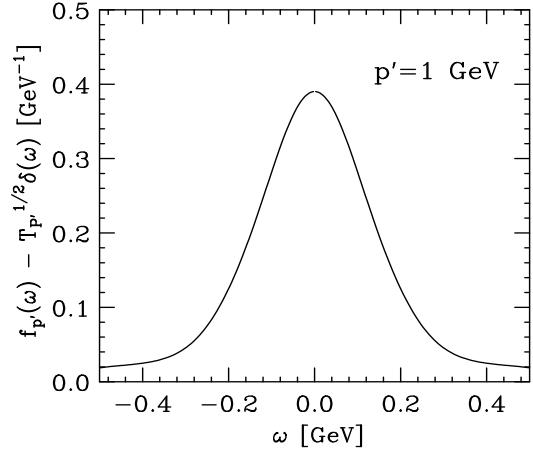
,angle=00,width=7.0cm,height=6.0cm}}
\vspace*{-.2in}
\caption{\small Finite width contribution to the folding function 
of Eq.(\protect\ref{ff}), evaluated
for a 1 GeV nucleon knocked out from oxygen using a parametrization of the 
measured NN scattering amplitude \protect\cite{tomon} and the wave function
of ref. \protect\cite{16Owf}.}
\label{folding}
\end{figure}

Fig. \ref{folding} shows the quantity 
$f_{{\bf p}^\prime}(\omega) - \delta(\omega) T_{{\bf p}^\prime}^{1/2}$, 
evaluated 
for a 1 GeV nucleon knocked out from oxygen using  a parametrization 
of the measured NN scatering amplitude \cite{tomon} and the wave function 
of ref. \cite{16Owf}.
%%%%%%%%%%%%%%%%%%%%%%%%%%%%%%%%%%%%%%%%%%%%%%%%%%%%%%%%%%%%%%%%%%%%%%%%%%%%
%%%%%%%%%%%%%%%%%%%%%%%%%%%%%%%%%%%%%%%%%%%%%%%%%%%%%%%%%%%%%%%%%%%%%%%%%%%%%%

\begin{figure}[hbt]
\centerline{\psfig{figure=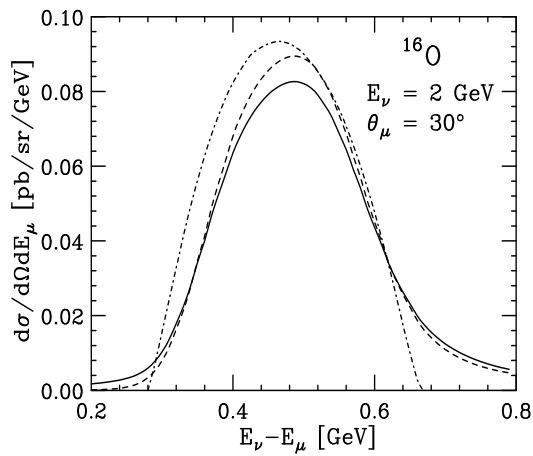
,angle=00,width=7.0cm,height=6.0cm}}
\vspace*{-.2in}
\caption{\small Cross section of the process $\nu + ^{16}O \rightarrow \mu 
+ p + X$. Solid line: Full calculation, including FSI.
Dashed line: IA calculation. Dot-dash line: Fermi gas model with
$k_F =$ 220 MeV and $\overline \epsilon$ = 25 MeV.}
\label{xsec:2}
\end{figure}

The effect of FSI on the calculated cross section of the process
$\nu_\mu + ^{16}O \rightarrow \mu^- + p + X$ at neutrino energy $E_\nu =$ 2 GeV
and muon scattering angle 30$^\circ$ is illustrated in Fig. \ref{xsec:2}.
It clearly appears that, in spite of the fact that the finite width component
contributes only $\sim$ 15 \% of the folding function normalization, the
inclusion of FSI produces a sizable redistribution of the strength, resulting 
in a quenching of the peak and an enhancement of the tails.

%%%%%%%%%%%%%%%%%%%%%%%%%%%%%%%%%%%%%%%%%%%%%%%%%%%%%%%%%%%%%%%%%%%%%%%%%%%%
%%%%%%%%%%%%%%%%%%%%%%%%%%%%%%%%%%%%%%%%%%%%%%%%%%%%%%%%%%%%%%%%%%%%%%%%%%%%%%

\section{Summary and conclusions}

Nuclear many-body theory provides a consistent framework 
allowing for a description of the continuum nuclear response
that takes into account dynamical effects not included in the 
FG model.

The approach discussed in this paper, that has been extensively 
applied to the analysis of electron-nucleus scattering data, is based on 
the IA picture, in which the lepton probe scatters off individual nucleons 
distributed in momentum and energy according to the spectral function
$P({\bf p},E)$. 

While this picture proved to work exceedingly well for scattering of few GeV
electrons, its applicability in the region of lower energies relevant to 
neutrino experiments needs to be quantitatively investigated. For example, 
the effect of Pauli blocking of the struck nucleon is likely to play
a non negligible role at $E_\nu \lsim$ 700 MeV.

Comparison between the results of the FG model and those
obtained using a realistic spectral function shows that the 
inclusion of dynamical correlation effects, leading to the appearance of 
high momentum and high removal energy components, sizably affects the 
response.

Corrections to the IA cross sections due to the occurrence of FSI,  
which have long been recognized to play a major role
in redistributing the inclusive strength, can also be taken into account 
in a consistent fashion within nuclear many-body theory.

As a final remark, it s worthwhile mentioning that the proposed many-body 
approach 
can be readily generalized to include the contribution of 
inelastic channels in a fully consistent fashion.\\ \\

\noindent{\bf Acknowledgments}\\

The author would like to thank Steven Pieper for providing the Monte
Carlo configurations employed in the calculation of FSI effects. 
Many illuminating discussions with Adelchi Fabrocini, Stefano Fantoni, 
Vijay Pandharipande and Ingo Sick are also gratefully acknowledged.

\end{document}